\documentclass[final,5p,twocolumn,12pt]{elsarticle}
\usepackage{graphics}

\journal{Physics Letters B}

\begin{document}

\begin{frontmatter}

\title{Antihydrogen Formation Dynamics in a Multipolar Neutral Anti-atom Trap} 

\author[aarhus]{G.B. Andresen}
\author[swan]{W. Bertsche}
\author[aarhus]{P.D. Bowe}
\author[berk]{C. Bray}
\author[swan]{E. Butler}
\author[rio]{C.L. Cesar}
\author[berk]{S. Chapman}
\author[swan]{M. Charlton}
\author[berk]{J. Fajans}
\author[triumf]{M.C. Fujiwara}
\author[triumf]{D.R. Gill}
\author[aarhus]{J.S. Hangst}
\author[ubc]{W.N. Hardy}
\author[tokyo]{R.S. Hayano}
\author[fraser]{M.E. Hayden}
\author[swan]{A.J. Humphries}
\author[calg]{R. Hydomako}
\author[swan]{L.V. J\o rgensen}
\author[swan]{S.J. Kerrigan}
\author[triumf]{L. Kurchaninov}

\author[rio]{R. Lambo}
\author[swan]{N.~Madsen}
\author[liver]{P. Nolan}
\author[triumf]{K. Olchanski}
\author[triumf]{A. Olin}
\author[berk]{A. Povilus}
\author[liver]{P. Pusa}
\author[auburn]{F. Robicheaux}
\author[nrcn]{E. Sarid}
\author[ubc]{S. Seif El Nasr}
\author[tokyo,riken]{D.M. Silveira}
\author[triumf]{J.W. Storey}
\author[calg]{R.I. Thompson}
\author[swan]{D.P. van der Werf}
\author[berk]{J.S. Wurtele}
\author[riken]{Y. Yamazaki}


\address[aarhus]{Department of Physics and Astronomy, Aarhus University, 8000 Aarhus C, Denmark}
\address[swan]{Department of Physics, Swansea University, Swansea SA2 8PP, United Kingdom}
\address[berk]{Department of Physics, University of California at Berkeley, Berkeley, CA 94720-7300, USA}
\address[rio]{Instituto de F{\'i}sica, Universidade Federal do Rio de Janeiro, Rio de Janeiro 21941-972, Brazil}
\address[triumf]{TRIUMF, 4004 Wesbrook Mall Vancouver, BC V6T 2A3, Canada}
\address[ubc]{Department of Physics and Astronomy, University of British Columbia, Vancouver, BC V6T 1Z1, Canada}
\address[tokyo]{Department of Physics, University of Tokyo, Tokyo 113-0033, Japan}
\address[fraser]{Department of Physics, Simon Fraser University, Burnaby, BC V5A 1S6 Canada}
\address[calg]{Department of Physics and Astronomy, University of Calgary, Calgary T2N 1N4, Canada}
\address[liver]{Department of Physics, University of Liverpool, Liverpool L69 7ZE, United Kingdom}
\address[auburn]{Department of Physics, Auburn University, Auburn, AL 36849-5311, USA}
\address[nrcn]{Department of Physics, NRCN-Nuclear Research Center Negev, Beer Sheva, IL-84190, Israel}
\address[riken]{Atomic Physics Laboratory, RIKEN, Saitama 351-0198, Japan}

\author{ALPHA Collaboration}


\begin{abstract} 
Antihydrogen production in a neutral atom trap formed by an octupole-based magnetic field minimum is demonstrated using field-ionization of weakly bound anti-atoms. Using our unique annihilation imaging detector, we correlate antihydrogen detection by imaging and by field-ionization for the first time. We further establish how field-ionization causes radial redistribution of the antiprotons during antihydrogen formation and use this effect for the first simultaneous measurements of strongly and weakly bound antihydrogen atoms. Distinguishing between these provides critical information needed in the process of optimizing for trappable antihydrogen. These observations are of crucial importance to the ultimate goal of performing CPT tests involving antihydrogen, which likely depends upon trapping the anti-atom.
\end{abstract}


\end{frontmatter}

Antihydrogen atoms ($\bar{\mbox{H}}$) are of fundamental interest due to the promise of sensitive tests of CPT symmetry based on comparisons of the spectra of hydrogen and antihydrogen. Cold $\bar{\mbox{H}}$  was first synthesized by the ATHENA collaboration \cite{amoretti2002} at the CERN Antiproton Decelerator (AD) \cite{belochitskii2001} in 2002 and subsequently by the ATRAP collaboration \cite{gabrielse2002}.
In these, and all later experiments, the neutral $\bar{\mbox{H}}$, which were produced in Penning traps from cold plasmas of positrons ($\mbox{e}^+$) and antiprotons ($\bar{\mbox{p}}$), escaped the production volume, either to annihilate or to be field-ionized. For future experiments on $\bar{\mbox{H}}$, it is highly desirable, and possibly necessary, to be able to trap and hold the neutral anti-atoms.  

In this letter we demonstrate the first $\bar{\mbox{H}}$ formation in an octupole-based magnetic minimum neutral atom trap and, for the first time, correlate the $\bar{\mbox{H}}$ detection by field-ionization and by annihilation imaging. We observe a decrease in the number of $\bar{\mbox{H}}$ formed as the atom trap depth is increased. Using detailed plasma and annihilation diagnostics, we present new insights into how field-ionization influences $\bar{\mbox{p}}$ transport during $\bar{\mbox{H}}$ production both with and without the neutral atom trap. We use this transport as a sensitive diagnostic of weakly bound, field-ionizable antihydrogen, and make the first simultaneous measurement of strongly and weakly bound states. We discuss how distinguishing between these is important for optimizing production of trappable $\bar{\mbox{H}}$. These studies were performed during the 2008 AD beamtime over a period of 4-6 weeks of the 24 week total.

The ALPHA apparatus used for the experiments presented here is designed to hold $\bar{\mbox{H}}$ in an octupole-based magnetic field minimum trap \cite{bertsche2006} superposed on a charged particle trap, and has been described in detail elsewhere \cite{andresen2008a}. Our charged particle traps are of the Penning-Malmberg type, where a uniform solenoidal field ensures radial confinement, whilst electric fields provide axial confinement.  These traps are cooled to $\sim$~7.5~K by the same cryostat used to cool the superconducting magnets that provide the fields for the minimum-B trap. 

The addition of  a transverse multipole B-field to a Penning trap limits the allowed radial extent of the trapped plasmas in that it induces a critical radius ($r_{crit}$) beyond which charged particles are lost \cite{fajans2005,fajans2008}. This arises because the confined low energy particles follow magnetic field-lines, and these intersect with the trap wall on introduction of a transverse multipole field. Even at lower radii, the azimuthally asymmetric multipole fields may perturb the plasma, leading to expansion and heating \cite{fajans2005,amoretti2006} similar to the effects of other static field trap asymmetries \cite{notte1994}. To minimize the influence of these fields on the charged particles used for $\bar{\mbox{H}}$ formation, we use an octupole to provide the transverse minimum-B, rather than a quadrupole as in the prototypical Ioffe-Pritchard geometry \cite{pritchard1983}. In order to further reduce transverse-field effects we have developed techniques to characterize and reduce the radial extent of our various particle species \cite{andresen2008b,andresen2008c}. 

The AD delivers $\sim$3$\times$10$^7$ $\bar{\mbox{p}}$ every 100~s. We slow these via passage through $\sim$218~$\mu$m of aluminium foil and trap a fraction of them. The trapped $\bar{\mbox{p}}$ then cool through collisions with a pre-loaded electron plasma containing $\sim$2$\times$10$^7$ particles, which cool through synchrotron radiation. This trapping and cooling is carried out in a 3~T field, formed by an exterior solenoid permanently at 1~T and a variable inner solenoid held at 2~T \cite{andresen2008a}. The high field increases our trapping efficiency and cooling rate \cite{andresen2008a}. We typically stack up to eight shots of $\bar{\mbox{p}}$ from the AD, resulting in about 2$\times$10$^5$ cold $\bar{\mbox{p}}$. These are then radially compressed, using a technique \cite{andresen2008c} based upon rotating-wall compression of $\mbox{e}^+$ and $\mbox{e}^-$ plasmas \cite{huang1997,greaves2000,jorgensen2005,funakoshi2007}, in preparation for mixing with $\mbox{e}^+$. Subsequently, the internal solenoid is ramped to zero, and the $\bar{\mbox{p}}$ are moved to the mixing region. A lower axial B-field allows a deeper trap for the neutral anti-atoms. The $\bar{\mbox{p}}$ plasma has a radius of 1.0~mm at 1~T in the mixing region, as measured using our Micro Channel Plate-based diagnostic (MCP) \cite{andresen2008c}.  

Positrons are accumulated from a $^{22}$Na source using N$_2$ buffer gas for capture and cooling \cite{surko1992}. We typically use 7$\times$10$^7$ $\mbox{e}^+$, which we accumulate in about 200~s. The $\mbox{e}^+$ are transferred to the mixing region \cite{jorgensen2005} and after radial compression, form a plasma of radius $\sim$1.5~mm and density $\sim$7$\times$10$^8$~cm$^{-3}$. 

\begin{figure}[hbt]
\centerline{\resizebox{\columnwidth}{!}{\includegraphics{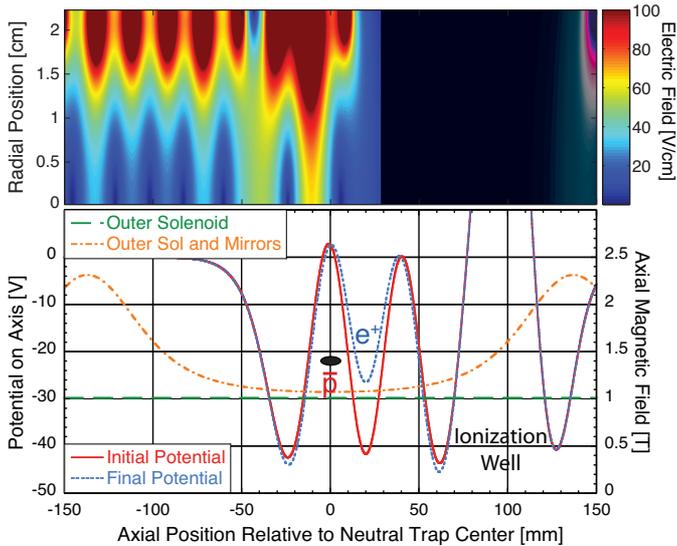}}}
\caption{Plots of the electric and magnetic fields used to create $\bar{\mbox{H}}$ in a minimum-B neutral atom trap. The lower plot shows the axial magnetic fields with (dot-dash) and without (long dash) the magnetic mirrors and the electric potential on axis at the beginning (solid line) and the end (short dash) of the mixing cycle. The black ellipse indicates the initial energy of the $\bar{\mbox{p}}$. The top plot shows the corresponding total electric field strength versus radius and axial position at the end of the mixing. }
\label{fig1}
\end{figure}

$\bar{\mbox{H}}$ is created using the modification proposed in Ref. \cite{madsen2006} to the most common mixing scheme  and is similar to one recently used \cite{gabrielse2008} and simulated \cite{ordonez2008}. This typically involves mixing of the $\mbox{e}^+$ and $\bar{\mbox{p}}$ in a variant of the nested Penning trap \cite{gabrielse1988}, which enables the oppositely charged species to be held in close proximity, and facilitates the injection of one into the other. After ramping up the neutral atom trap (the octupole and the two co-axial mirror coils which form the magnetic minimum), which takes about 30 s, $\bar{\mbox{p}}$ are injected into their well with non-zero axial energy, low enough to avoid contact with the $\mbox{e}^+$. The $\mbox{e}^+$ and the $\bar{\mbox{p}}$ are held in adjacent potential wells (Figure \ref{fig1}). By slowly raising the voltage that confines the $\mbox{e}^+$, over a period of typically 100 s, we bring the two particle species into contact and form $\bar{\mbox{H}}$. In this way $\bar{\mbox{p}}$ are brought into contact with $\mbox{e}^+$ at very low axial energy, with the aim of producing $\bar{\mbox{H}}$ with low kinetic energy. At the end of the voltage ramp (see Figure \ref{fig1}) all particles are ejected for counting. 

During mixing, the $\bar{\mbox{p}}$ can have a number of different fates. 1) A $\bar{\mbox{p}}$ can form neutral $\bar{\mbox{H}}$, which, if not magnetically trapped, will only be modestly influenced by the magnetic fields, and thus move approximately in a straight line from its origin to the wall, on which it annihilates \cite{madsen2005}. We will refer to these as strongly bound $\bar{\mbox{H}}$ as they survive the electric fields of the trap. The highest fields in the trap are of order 100~V cm$^{-1}$, so strongly bound $\bar{\mbox{H}}$ corresponds to binding energy greater than $\sim$7.5~meV. 2) The $\bar{\mbox{p}}$ forms $\bar{\mbox{H}}$, which is in turn field-ionized. Depending on where in the volume this happens, this $\bar{\mbox{p}}$ may end up being re-trapped or lost to annihilation. We will refer to these as weakly bound $\bar{\mbox{H}}$, or bound by less than $\sim$7.5~meV. 3) A $\bar{\mbox{p}}$ or $\bar{\mbox{H}}$ can annihilate with a residual gas atom or ion. 4) A $\bar{\mbox{p}}$ can be lost due to radial transport out of the trap without ever forming $\bar{\mbox{H}}$. Such transport is common in Penning trap experiments and is usually enhanced in the presence of field-inhomogeneities. These so-called $\bar{\mbox{p}}$-only losses result in localized annihilation "hot-spots" on the wall \cite{fujiwara2004}.  

We employ two techniques to detect $\bar{\mbox{p}}$ and $\bar{\mbox{H}}$. The most sensitive uses a silicon vertex detector, which reconstructs the tracks of charged pions from $\bar{\mbox{p}}$ annihilations, thereby locating the annihilation vertices \cite{amoretti2002, andresen2009}. As was demonstrated by ATHENA (with no neutral atom trap), the $\bar{\mbox{p}}$-only "hot-spots" \cite{madsen2005,fujiwara2004} are to be contrasted with the $\bar{\mbox{H}}$ annihilations, which produce a smooth and radially symmetric vertex distribution \cite{madsen2005}. The vertex detector used here has a position resolution of about 5~mm (one sigma) \cite{andresen2009}, due primarily to the uncertainty in reconstructing the $\bar{\mbox{p}}$ vertex as a result of scattering of the pions in the material between the annihilation point and the detector. As only about 20\% of the recorded annihilations are reconstructed in the present condition, we can also use a simple trigger, which requires a minimum of two triggered silicon modules, as a proxy for $\bar{\mbox{H}}$ formation, as applied previously \cite{amoretti2004a}. From Monte Carlo simulations, we estimate that this trigger has an efficiency of about 95\% for $\bar{\mbox{p}}$ annihilation events. However, to determine which fraction of $\bar{\mbox{p}}$ annihilations are due to $\bar{\mbox{H}}$, the full vertex distribution must be analysed (more below). A second, complementary, method to establish $\bar{\mbox{H}}$ formation is to intentionally field-ionize weakly bound $\bar{\mbox{H}}$, trap their $\bar{\mbox{p}}$ in a particular well and then count the number held \cite{gabrielse2002,gabrielse2008}. This was done by deliberately ejecting them from the well onto the aluminum degrader foil and monitoring the resulting annihilations with external detectors. This so-called ionization well is shown in Figure \ref{fig1}, and we refer to the $\bar{\mbox{p}}$ in this well as intentionally field-ionized $\bar{\mbox{H}}$.

\begin{figure}[hbt]
\centerline{\resizebox{\columnwidth}{!}{\includegraphics{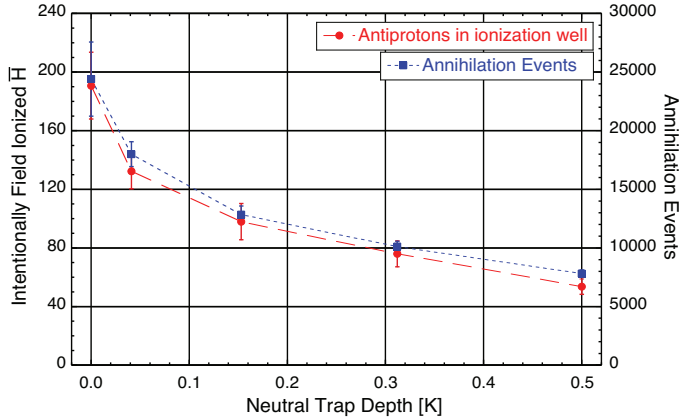}}}
\caption{Number of intentionally field-ionized $\bar{\mbox{H}}$, detected following ejection (see text), and annihilation events as a function of the depth of the neutral atom trap for ground state $\bar{\mbox{H}}$. The depth of the well in Tesla can be found by dividing the ordinate values by 0.67 (see e.g. \cite{holzscheiter2004}). The uncertainties represent variations in  reproducibility. All values are normalized to 10$^5$ $\bar{\mbox{p}}$ brought into mixing. The scaling accidentally makes the numbers overlap at zero field.}
\label{fig2}
\end{figure}

Figure \ref{fig2} shows the number of annihilation events and the number of $\bar{\mbox{H}}$ intentionally field-ionized during mixing at different depths of the neutral atom trap. Antihydrogen was formed by merging $\mbox{e}^+$ with $\bar{\mbox{p}}$ for 100~s. Without $\mbox{e}^+$, no field-ionization or annihilation was observed. With $\mbox{e}^+$, we see  evidence of intentionally field-ionized $\bar{\mbox{H}}$ at all neutral atom trap depths. We observe that the number of $\bar{\mbox{H}}$ thus detected decreases with trap depth. This drop could be caused by an increase in the (unmeasured) plasma temperature induced by the transverse magnetic fields, since field inhomogeneities are known to cause plasma expansion and heating \cite{fajans2005,amoretti2006}. The annihilation events show a trend similar to that of the intentionally field-ionized $\bar{\mbox{H}}$.

Conducting the experiment with only the mirror coils of the neutral atom trap gives the same mixing results as that with no trap. However, with only the octupole field, the results are equivalent to mixing in the full neutral atom trap, implying that it is the octupole which is responsible for the drop in $\bar{\mbox{H}}$ formation in both cases. These results differ from an earlier report from ATRAP \cite{gabrielse2008} where an increase in $\bar{\mbox{H}}$ formation with neutral atom trap depth was observed. However, the ATRAP increase was also present without their transverse (quadrupole) fields. A likely explanation for this is that since the ATRAP mirror coils are closer together than in our experiment, their axial B-field in the formation region increases significantly (from 1.0~T to 2.2~T) when their trapping field is turned on. In ALPHA the mirrors add only 0.1~T to the 1.0~T main solenoidal field in the $\bar{\mbox{H}}$ formation region. Increasing the axial field will increase the $\mbox{e}^+$ synchrotron cooling rate and can increase the plasma density. Both higher densities and lower temperatures were observed by ATHENA to increase formation rates \cite{funakoshi2007,amoretti2004c}.   

\begin{figure}[hbt]
\centerline{\resizebox{\columnwidth}{!}{\includegraphics{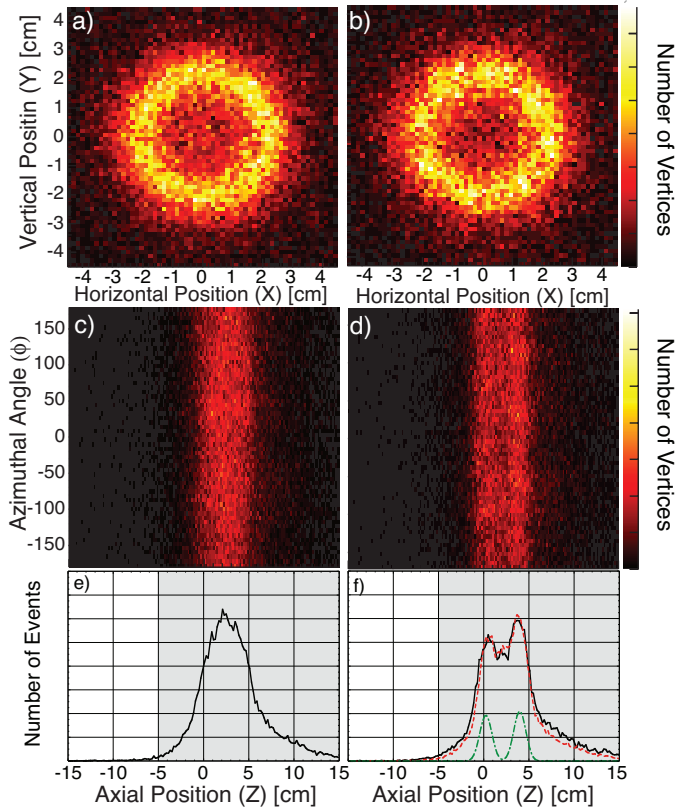}}}
\caption{(a-b) Azimuthal projections of the $\bar{\mbox{p}}$ annihilation vertex distributions during mixing with (a) no neutral atom trap and (b) the full trap. (c-d) Corresponding z-$\phi$ distributions. (e-f) Corresponding axial (z) distributions. (f) Dashed (red): Fit to the distribution, see text; Dot-dashed (green) peaks in fit. The shaded area marks the three layer part of the detector used for tracking. Left of this area the detector has only one layer of silicon. The slight asymmetry in the axial distributions (in particular the tails) is due to the lower reconstruction efficiency outside the three layer section. For clarity the plots have been normalized to have the same total number of events. The zero axial position is the center of the neutral atom trap.}
\label{fig3}
\end{figure}

To investigate the correlation between annihilation events and intentional field-ioniza\-tion counts, we next consider the annihilation vertex distribution. Figure \ref{fig3} shows the projection of the full three-dimensional data onto the azimuthal plane during mixing from the experiments (a) with no neutral atom trap (mirrors and octupole off) and (b) with our maximum trap depth of 0.5~K, corresponding to an octupole field at the wall of 1.4~T. The observed ring structure in the projections is consistent with the electrode diameter of $\sim$44.6~mm. The "no neutral atom trap" measurement shows a smooth, radially symmetric vertex distribution and no hot-spots (Figure \ref{fig3}a,c,e). Consequently, we infer that all the annihilation events in that experiment stem from $\bar{\mbox{H}}$ annihilations similar to those observed by ATHENA \cite{madsen2005}. From the data (Figure \ref{fig2}) we find that (24$\pm$3)\% of the $\bar{\mbox{p}}$ form $\bar{\mbox{H}}$ that impacts on the wall (i.e. strongly bound), whilst (0.19$\pm$0.02)\% produce $\bar{\mbox{H}}$ that is intentionally field-ionized (i.e. weakly bound). These figures were derived by normalizing the counted events in each case to 10$^5$ $\bar{\mbox{p}}$ taking part in each mixing experiment.

In the presence of the full neutral atom trap, the fraction of annihilations on the wall and the fraction of intentionally field-ionized $\bar{\mbox{H}}$ have both dropped by 2/3 to (7.8$\pm$0.3)\% and (0.054$\pm$0.005)\% respectively. However, the vertex distributions show some indications of anisotropy. In particular, there are evident side peaks in the axial distribution (Fig. \ref{fig3}f). It is beyond the scope of this letter to characterize the radial imaging with the neutral atom trap fields, however, the axial distribution can provide some insights into the particle dynamics during mixing, and help elucidate what fraction of the annihilations on the wall is due to $\bar{\mbox{H}}$. 

In the presence of the octupole, the axial vertex distribution can depend on the $\bar{\mbox{p}}$ radial density distribution because, as mentioned earlier, a Penning-trap with a superposed multipole field has, for a given axial excursion, a critical radius  beyond which particles are lost \cite{fajans2008}. We can use the octupole as a radial diagnostic and measure the radial distribution of the $\bar{\mbox{p}}$ by ramping up the octupole field whilst monitoring $\bar{\mbox{p}}$ annihilations \cite{andresen2008b}. In the experiments discussed in this article the smallest $r_{crit}$ for trapped particles is 9~mm. This is much larger than the pre-mixing radii of any of our plasmas, thus the neutral atom trap cannot initially induce this type of loss.  For the experiments with the neutral atom trap described above, the octupole and mirrors are ramped before mixing, and no $\bar{\mbox{p}}$ losses are observed during the ramp. If, however, the magnets are energized {\sl after} completion of the mixing cycle, losses are observed during the ramp. We find that (6.1$\pm$0.2)\% of the $\bar{\mbox{p}}$ annihilate during this octupole ramp. The remaining cloud of $\bar{\mbox{p}}$ has a radius smaller than $r_{crit}$. After $\bar{\mbox{H}}$ formation, we find, using our MCP diagnostic, that $\sim$69\% of the $\bar{\mbox{p}}$ injected are left  with the same radial extent as before injection. Without $\mbox{e}^+$, we observe no losses or change in radial distribution. We find, using analysis similar to that in Ref. \cite{fajans2008}, that the radially redistributed $\bar{\mbox{p}}$ have a constant density from $r_{crit}$ out to the wall.  (The $\bar{\mbox{p}}$ density beyond $r_{crit}$ is determined by counting the antiproton annihilation events whilst ramping the octupole. At any given moment during the ramp the octupole field is known, and thus $r_{crit}$ is determined. To obtain the radial distribution the time distribution of annihilations is mapped during the ramp to a distribution of annihilations as a function of $r_{crit}$, as described in detail in \cite{fajans2008}.) We conclude that $\bar{\mbox{H}}$ formation causes some $\bar{\mbox{p}}$ to be radially transported much beyond either of the initial plasma radii while remaining trapped.

We attribute the radially redistributed $\bar{\mbox{p}}$ to weakly bound $\bar{\mbox{H}}$ which are field-ionized at high radius where the E-fields are large (Figure \ref{fig1}). Recall that no radial redistribution is observed without $\bar{\mbox{H}}$ formation. The intentional field-ionization well has a solid angle with respect to the center of the $\mbox{e}^+$ of between 1.8\% and 5\%, depending on the binding energy of the $\bar{\mbox{H}}$. For the same binding energies as those stripped in the ionization well on axis, the main $\bar{\mbox{p}}$ trapping potentials ionize over a total solid angle of $\sim$80\%. Thus, from simple geometrical considerations, scaling the intentional ionization well result of (0.19$\pm$0.02)\% leads us to expect that between 3\% and 8\% of the $\bar{\mbox{p}}$ create $\bar{\mbox{H}}$ that is field-ionized resulting in a re-trapped $\bar{\mbox{p}}$. If we assume that the redistributed $\bar{\mbox{p}}$ also have constant density in the region not covered by the octupole diagnostic, the (6.1$\pm$0.2)\% $\bar{\mbox{p}}$ redistributed to the region from $r_{crit}$ to the wall results in an estimated $\sim$7.3\% being field ionized and re-trapped in the full trap volume. This agrees well with the above estimate of 3-8\% and further supports the notion that the post-mixing high radius $\bar{\mbox{p}}$ originate from field-ionized $\bar{\mbox{H}}$. 

To further investigate this, we conducted a separate experiment with a lower density $\mbox{e}^+$ plasma with 7$\times$10$^7$ $\mbox{e}^+$ of radius 3.0~mm and density 3$\times$10$^8$~cm$^{-3}$. From the annihilation events we found that (12.2$\pm$1.6)\% of the $\bar{\mbox{p}}$ form strongly bound $\bar{\mbox{H}}$. This is about half the yield we found at the higher density, consistent with the decrease in yield with decrease in density observed in ATHENA \cite{funakoshi2007}. However, for weakly bound $\bar{\mbox{H}}$ we observed that the absolute number of intentionally field-ionized $\bar{\mbox{H}}$ {\sl increased} by (29$\pm$3)\% while the number from the octupole radial diagnostic {\sl decreased} by (30$\pm$2)\%. The difference between the two measures of field-ionized $\bar{\mbox{H}}$ is likely due to the differences in the fields experienced by $\bar{\mbox{H}}$ moving radially and axially away from the formation region. Taken together the difference in the changes of these three different measures of $\bar{\mbox{H}}$ could be caused by a change in the $\bar{\mbox{H}}$ binding energy distribution. 
$\bar{\mbox{H}}$ that is susceptible to field ionization in the ambient plasma and trap fields is unlikely to be trapped. Using the radial octupole diagnostic we can use the field-ionization induced radial transport as an efficient detector of weakly bound (field-ionizable) $\bar{\mbox{H}}$. The observation that the strongly and weakly bound populations do not simply scale when the $\mbox{e}^+$ density is changed is an indication that intentionally field-ionized $\bar{\mbox{H}}$ may be an insufficient indicator for the production of potentially trappable $\bar{\mbox{H}}$. By maximizing the ratio of $\bar{\mbox{H}}$ annihilating on the inner electrode wall (strongly bound) versus radially redistributed $\bar{\mbox{p}}$ (stemming from weakly bound $\bar{\mbox{H}}$) we can thus optimize for trappable $\bar{\mbox{H}}$.

Particles detected by the octupole radial diagnostic after mixing, would, independent of their origin, also have been lost if the octupole was already on, although the annihilation pattern would be different since the particles do not originate from the same trajectories when they experience loss. Thus, for mixing in the neutral atom trap, some fraction of the annihilation events must stem from octupole-induced loss of the radially redistributed $\bar{\mbox{p}}$. The $\bar{\mbox{p}}$ resulting from field-ionized $\bar{\mbox{H}}$ will follow magnetic field lines. If the ionization occurred at a sufficiently high radius, they will always follow the field lines into the wall, tending to produce, by the magnetic field symmetry, two peaks in the axial distribution, one on either side of the $\mbox{e}^+$ plasma. This is qualitatively as observed (Figure \ref{fig3}f). We conclude that field-ionization induced radial transport is the likely cause of the two side peaks in the axial distribution. 

We can estimate the fraction of $\bar{\mbox{H}}$ directly impacting the wall in the full neutral atom trap measurement by assuming that the axial annihilation distribution of strongly bound $\bar{\mbox{H}}$ remains unaltered by the neutral atom trap. We fit the neutral atom trap distribution (Figure \ref{fig3}f) with the sum of the no-trap distribution (Figure \ref{fig3}e) and two additional Gaussian peaks representing the annihilations from weakly bound $\bar{\mbox{H}}$  which ionizes beyond $r_{crit}$ (shown together as the dashed red curve in Figure \ref{fig3}f). The fit results in peaks of width $\sim$7~mm (dot-dashed green curve), which is larger than the axial resolution. From the fit we estimate that (83$\pm$5)\% of the events in the full neutral atom trap are due to strongly bound $\bar{\mbox{H}}$ hitting the wall directly, with the remainder originating from field ionized $\bar{\mbox{H}}$. 

As a consistency check of this model, we can again refer to the octupole radial diagnostic results. For this, we assume that the fraction of events on the wall relative to those caused by radial redistribution is independent of neutral atom trap depth. As the octupole diagnostic showed that (6.1$\pm$0.2)\% of the $\bar{\mbox{p}}$ redistribute to beyond $r_{crit}$, when (24$\pm$3)\% form $\bar{\mbox{H}}$ that reaches the wall, we find that 24/(30.1) = (80$\pm$2)\% of the events during a given mixing cycle in the full neutral atom trap originate from $\bar{\mbox{H}}$ annihilating on the wall. This is consistent with the fit result of (83$\pm$5)\%.  If $\sim$80\% of the annihilation events during mixing in the neutral atom trap are $\bar{\mbox{H}}$, the conversion efficiency of $\bar{\mbox{p}}$ into $\bar{\mbox{H}}$ that impacts directly on the wall is (6.2$\pm$0.3)\%.

In conclusion, we have demonstrated the first $\bar{\mbox{H}}$ formation in an octupole based neutral atom trap and, for the first time, correlated the results from field-ionization and imaging. We observe a drop in $\bar{\mbox{H}}$ formation with increased neutral atom trap depth, likely due to plasma expansion in the non-homogeneous magnetic fields. When mixing in the neutral atom trap, we further observe two axially separated peaks in an annihilation distribution otherwise consistent with $\bar{\mbox{H}}$. We argued that these are caused by an observed radial redistribution of $\bar{\mbox{p}}$ during mixing. We have shown that this redistribution is due to $\bar{\mbox{H}}$ unintentionally ionized on the electric fields of our charged particle traps. By combining field-ionization measurements and annihilation detection we can distinguish between weakly and strongly bound $\bar{\mbox{H}}$. We can therefore selectively optimize for strongly bound, potentially trappable, $\bar{\mbox{H}}$.

This work was supported by CNPq, FINEP (Brazil), NSERC, NRC/TRIUMF (Canada), FNU (Denmark), ISF (Israel), MEXT (Japan), The Leverhulme Trust, EPSRC (UK) and DOE (USA).

\end{document}